\documentclass[twocolumn,showpacs,preprintnumbers,amsmath,amssymb]{revtex4}

\begin{document}
\title{Direct measurement of three-body interactions}
\author{Matthias Brunner, Jure Dobnikar, Hans-Hennig von Gr\"unberg, and Clemens Bechinger}
\address{University of Konstanz, Physics Department, 78457 Konstanz, Germany} 

\begin{abstract}
  Three-body interactions have been measured among three charged
  colloidal particles in deionized solvent. Two of the particles have
  been confined to an optical line-trap while the third one was
  approached by means of a focused laser beam. The experimentally
  determined three-body interactions are attractive and roughly of
  the same magnitude and range as the pair-interactions.  In addition,
  numerical calculations have been performed, which show good
  agreement with the experimental results.
\end{abstract}
\maketitle

Precise knowledge of particle interactions is of utmost importance for
the understanding of thermodynamic properties of condensed matter. We
typically treat interactions in a pair-wise fashion, but if the
governing physical equations are non-linear, interactions between two
particles can be modified by a close third or fourth particle. The
total energy is then no longer given by the sum of all pair-potentials
but additional many-body interactions appear.  Physical examples of
many-body interactions are abundant: inter atomic potentials, electron
screening in metals \cite{Hafner}, island formation on surfaces
\cite{Osterlund}, chemical processes in solids \cite{Ovchinnikov} and
even noble gases which posses a closed-shell electronic structure
\cite{Axilrod,Jakse}. In view of its general importance it seems
surprising that until now no direct measurement of many-body
interactions has been performed, however, it is hard to imagine that
such a measurement could be performed in an atomic system, where
positional information is provided in an integrated form, i.e. by
structure factors or pair-correlation functions. Many-body
interactions can be directly evaluated only when the positional
information is provided, i.e. the particles' trajectories in space and
time are known.\\

In contrast to atomic systems, length- and time-scales in colloidal
systems are accessible with optical experiments and it is possible to
obtain individual particle trajectories.  Additionally, colloidal
interactions can be tuned over a wide range, simply by changing the
salt concentration (in contrast to atoms where interactions are
unchangeably dictated by their electronic structure). At sufficiently
small salt concentrations, the interaction range can reach several
$\mu$m.  If more than two colloids are within this range, a simple
pair-wise description breaks down and many-body interactions occur.
Indeed, three-body effects have been found in recent computer
simulations \cite{Loewen,Bratko,Carsten}. It was also demonstrated by
experiments \cite{gr} and simulations \cite{EPL,JCP}, that the
effective pair-potential of a two-dimensional colloidal liquid shows a
density dependence, hinting towards many-body effects.  Accordingly,
colloidal suspensions represent an ideal model system for systematic
investigations in this
field.\\

Here we present the first direct measurement of three-body
interactions in a suspension of charged colloidal particles. Two of
the particles were confined by means of a scanned laser tweezer to a
quasi-static line-shaped optical trap where they diffused due to
thermal fluctuations.  A third particle was localized in a point-like
laser tweezer at distance $d$ (see inset of Fig.1). When the third
particle was approached to the line trap, significant deviations from
pairwise additivity have been observed. This experimental finding is
also supported by the additionally performed Poisson-Boltzmann calculations.\\

We used a highly diluted aqueous suspension of charge-stabilized
silica spheres ($990$nm diameter), which has been confined in a silica
glass cuvette with $200 \mu$m spacing. The cuvette was connected to a
standard deionization circuit described elsewhere \cite{Palberg}.
Before each measurement the water was pumped through the ion exchanger
and typical ionic conductivities below $0.07 \mu$S/cm were obtained.
After the suspension was injected, the cuvette was disconnected from
the circuit during the measurements. This
procedure yielded stable and reproducible ionic conditions during the experiments.\\

Particle interaction measurements performed with scanned optical
tweezers have been reported by several other groups
\cite{Owen,Crocker}, therefore the technique is here only briefly
described. The focussed beam of an Ar$^{+}$ laser ($488$nm) was
scanned across our sample cell by means of a galvanostatically driven
mirror with a frequency of about 350 Hz. This yielded a Gaussian
intensity distribution along and perpendicular to the scanning
direction with halfwidths $ \sigma_{x} \approx 4.5 \mu$m and $
\sigma_{y} \approx 0.5 \mu$m, respectively. Due to the negatively
charged silica substrate, the particles experience a repulsive
vertical force balanced by the particle weight and the vertical
component of the light force. The potential in the vertical direction
is much steeper than the in-plane laser potential, therefore vertical
particle fluctuations can be disregarded. The particles were imaged
with a long-distance, high numerical aperture microscope objective
onto a CCD camera and the lateral positions of the particle centers
were determined with a resolution of about 25 nm by a particle
recognition algorithm.\\

We first inserted a single particle into the trap where it diffused
due to thermal fluctuations. From the positional probability
distribution $P(x,y)$, the laser potential $u _{L}(x,y)$ is directly
obtained via the Boltzmann distribution $P(x,y) \propto e^{- \beta u
  _{L}(x,y)}$, where $\beta^{-1}$ corresponds to the thermal energy of
the suspension. Next, we inserted a second particle in the trap. The
four-dimensional probability distribution is now $P(x _{1} , y _{1}, x
_{2}, y _{2}) = P _{12} e^{\beta (u_{L}(x _{1},y _{1}) + u _{L}(x
  _{2}, y _{2}) + U(r))}$ with $x _{i}$, $y _{i}$ being the position
of the i-th particle relative to the laser potential minimum and
$U(r)$ the distance dependent pair-interaction potential between the
particles. Since it is reasonable to assume that the pair interaction
depends only on the particle distance, we projected $P(x _{1} , y
_{1}, x _{2}, y _{2})$ onto a one-dimensional distance distribution
$P(r)$.  From the measured $P(r)$ we obtained the total potential
energy of the particles and after subtracting the external potential
we were left with the pair interaction potential $U(r)$.\\

The pair-interaction potential of two charge-stabilized particles in
the bulk is theoretically predicted \cite{Landau,Overbeek} to
correspond to a Yukawa potential

\begin{equation} \label{1}
\beta U(r) \equiv \beta u_{12}(r) = (Z^*)^{2} \lambda _{B} \left(
\frac{exp(\kappa R)}{1+\kappa R} \right)^2 \frac{exp(-\kappa r)}{r}
\end{equation}

where $Z^*$ is the renormalized charge \cite{Belloni} of the
particles, $\lambda _{B}$ the Bjerrum-length ( in water at room
temperature), $\kappa ^{-1}$ the Debye screening length (given by the
salt concentration), $R$ the particle radius and $r$ the center-center
distance of the particles. Fig.2 shows the experimentally determined
pair-potential (symbols) together with a fit to Eq.(1) (solid line).
As can bee seen, our data are well described by Eq.(1). As fitting
parameters we obtained $Z^* \approx 6500$ electron charges and $
\kappa^{-1} \approx 470nm$, respectively. $Z^*$ is in good agreement
with the predicted value of the saturated effective charge of our
particles \cite{Zsat,Trizac} and $\kappa ^{-1}$ agrees reasonably with
the bulk salt concentration in our suspension as obtained from the
ionic conductivity. Given the additional presence of a charged
substrate, it might seem surprising that Eq.(1) describes our data
successfully.  However, it has been demonstrated experimentally
\cite{Grier} and theoretically \cite{Stillinger,Netz} that a
Yukawa-potential captures the leading order interaction also for
colloids close to a charged wall. A single confining wall introduces
only a very weak (below $0.1 k _{B} T$) correction due to additional
dipole repulsion which is below our experimental resolution. Repeating
two-body measurements with different laser intensities (50mW to 600mW)
yielded within our experimental resolution identical pair potential
parameters. This also demonstrates that possible light-induced
particle interactions (e.g. optical binding \cite{Burns}) are neglegible. \\

Finally, we approached a third particle by means of an additional
laser trap at distance $d$ (cf. Fig.1) where it was localized during
the whole measurement. We carefully checked that the emply laser trap
(i.e. without the third particle) has no influence on the
pair-interaction potential on the particles in the line tweezer. From
the distance distribution of the two particles in the laser trap, we
can, applying the same procedure as in the two particle measurement,
extract the total interaction energy which is now also characterized
by the distance $d$ , i.e.  $U=U(r;d)$. Following the definition of
McMillan and Mayer \cite{McM}, the total interaction energy for three
particles $U(r;d)$ can be written as

\begin{equation} \label{2}
U(r;d) = u _{12}(r _{12}) + u _{13}(r _{13}) + u _{23}(r _{23}) + u
_{123}(r _{12}, r _{13}, r_{23})
\end{equation}

with $u _{ij}$ being the pair-potential between particles $i$ and $j$
as shown in Fig.2 and $u _{123}$ the three-body interaction potential.
$r _{12}$, $r _{23}$ and $r _{13}$ are the distances between the three
particles which can - due to the chosen symmetric configuration - be
expressed by the two variables $r = r _{12}$ and $d$. \\

The measured interaction energies $U(r;d)$ are plotted as symbols in
Fig.3 for several distances of the third particle ($d = 4.1, 3.1, 2.5,
1.6 \mu$m).  As expected, $U(r;d)$ becomes larger as $d$ decreases due
to the additional repulsion between the two particles in the trap and
the third particle. In order to test whether the interaction potential
can be understood in terms of a pure superposition of
pair-interactions, we first calculated $U(r;d)$ according to Eq.(2)
with $u _{123} = 0$. This was easily achieved because the positions of
all three particles were determined during the experiment and the
distance-dependent pair-potential is known from the two-particle
measurement described above (Fig.2). The results are plotted as dashed
lines in Fig.3. Considerable deviations from the experimental data can
be observed, in particular at smaller $d$. These deviations can only be
explained, if we take three-body interactions into account. Obviously,
at the largest distance, i.e.  $d = 4.1 \mu$m our data are well
described by a sum over pair-potentials which is not surprising, since
the third particle cannot influence the interaction between the other
two, if it is far enough from both.  In agreement with theoretical
predictions \cite{Carsten}, the three-body interactions therefore decrease
with increasing distance $d$.\\

According to Eq.2 the three-body interaction potential is simply given
by the difference between the measured $U(r,d)$ and the sum of the
pair-potentials (i.e. by the difference between the measured data and
their corresponding lines in Fig.3).  The results are plotted as
symbols in Fig.4.  It is seen, that $u _{123}$ is entirely attractive
and becomes stronger as the third particle is approached. The range of
$u _{123}$ is of the same order as the pair-interaction potentials.
To support our results, we also performed Poisson-Boltzmann (PB)
calculations, in a similar way as in \cite{Carsten}. The PB theory provides
a mean-field description in which the micro-ions in the solvent are
treated as a continuum, neglecting correlation effects between the
micro-ions.  It has repeatedly been demonstrated \cite{Groot,Levin} that in
case of monovalent micro-ions the PB theory provides a reliable
description of colloidal interactions.  We used the multi-centered
technique, described and tested in other studies \cite{EPL,JCP} to solve the
non-linear PB equation for the electrostatic mean-field potential
$\Phi$, which is related to the micro ionic charge density $\rho _{c}
= - ( \kappa^{2} / 4 \pi \lambda _{B}) \sinh \Phi $. Integrating the
stress tensor, depending on $\Phi$, over a surface enclosing one
particle, results in the force acting on this particle. Calculating
the force $f _{12}$ and from it the pair-potential between only two
particles, we first reproduced the measured pair-interaction in
Fig.2. The calculation of three-body potentials was then carried out
by calculating the total force acting on one particle in the line trap
(say, particle 1) and subtracting the corresponding pair-forces $f
_{12}$ and $f _{13}$ obtained previously in the two-particle
calculation.  The difference has been integrated to obtain the
three-body potential. The results are plotted as lines in Fig.4 and
show reasonable agreement with the experimental data (in particular
with respect to the range and size of $u _{123}$).  This strongly
supports our interpretation of the experimental results in terms of
three-body interactions.  The remaining deviations between theory and
experiment are probably due to small variations in salt concentration
but may also be due to small differences in the size and surface
charge of the colloidal particles used in the experiment, which have
been assumed to be identical in the PB calculations.\\

We have demonstrated that in case of three colloidal particles,
three-body interactions present a considerable contribution to the
total interaction energy and must therefore inevitably be taken into
account. Whenever dealing with systems comprised of many (much more
than three) particles, in principle also higher-order terms have to be
considered. We expect, however, that there is an intermediate density
regime, where the macroscopic properties of systems can be
successfully described by taking into account only two- and three-body
interactions. Indeed, there are systems where this was experimentally
observed \cite{Osterlund,Jakse}. At even larger particle densities
$n$-body terms with $n > 3$ have to be additionally considered, which may
partially compensate. Even in this regime, however, many-body effects
are not cancelled, but lead to notable effects, e.g. to a shift of the
melting line in colloidal suspensions, as recently demonstrated by PB
calculations \cite{EPL,JCP}.\\

Discussions with R. Klein, C. Russ and  E. Trizac and financial
support from the Deutsche Forschungsgemeinschaft (Grants Be1788/3 and
Gr1899) are acknowledged.

\section{Figure captions}

\noindent {\sl Figure 1} \quad Photograph of sample cell with two 
silica particles confined to a light trap created by an optical 
tweezers and a third particle trapped in a focused laser beam. The 
inset shows schematically the experimental geometry.

\noindent {\sl Figure 2} \quad Measured pair-interaction potentials 
$U(r)$ (symbols) in the absence of the third particle. The data agree
well with a Yukawa potential (solid line). In the inset the potential
is multiplied by $r$ and plotted logarithmically, so that Eq. 1
transforms into a straight line.

\noindent {\sl Figure 3} \quad Experimentally determined interaction 
energy $U(r)$ (symbols) for two particles in a line tweezers in the
presence of a fixed third particle with distance d on the
perpendicular  bisector of the line trap. For comparison the
superposition of three  pair-potentials is plotted as lines. Symbols
and lines are labeled by  the value of $d$.

\noindent {\sl Figure 4} Three-body potentials for different $d$. 
Measured three-body potentials indicated by symbols. The lines are
three-body potentials as obtained from the solutions of the nonlinear
Poisson-Boltzmann equation for three colloids arranged as in the
experiment. The parameters in the Poisson-Boltzmann calculation were
chosen so that the pair-interaction potentials were correctly
reproduced. Symbols and lines are labeled by the value of $d$.


\begin{thebibliography}{99}
\bibitem{Zsat} The saturated effective charge of our particles is about
6900 for water at room temperature.
\end{thebibliography}
\end{document}